\documentclass[]{spie}  

\usepackage{amsmath,amsfonts,amssymb}
\usepackage{graphicx}
\usepackage[colorlinks=true, allcolors=blue]{hyperref}

\title{Fast focal plane wavefront sensing on ground-based telescopes}

\author[a,b]{Benjamin L. Gerard}
\author[b,a]{Christian Marois}
\author[c]{Rapha\"el Galicher}
\author[b,a]{Jean-Pierre V\'eran}
\affil[a]{University of Victoria, Department of Physics and Astronomy, 3800 Finnerty Rd, 
Victoria, V8P 5C2, Canada}
\affil[b]{National Research Council of Canada, Astronomy \& Astrophysics Program \\
5071 West Saanich Rd\\
Victoria, V9E 2E7, Canada}
\affil[c]{Lesia, Observatoire de Paris, PSL Research University, CNRS, Sorbonne Universit\'es, Univ. Paris Diderot \\
UPMC Univ. Paris 06, Sorbonne Paris Cit\'e, 5 place Jules Janssen \\
92190 Meudon, France}

\authorinfo{Further author information: Send correspondence to Benjamin L. Gerard (bgerard@uvic.ca)}

\begin{document} 
\maketitle

\begin{abstract}
Exoplanet detection and characterization through extreme adaptive optics (ExAO) is a key science goal of future extremely large telescopes. This achievement, however, will be limited in sensitivity by both quasi-static wavefront errors and residual AO-corrected atmospheric wavefront errors. A solution to both of these problems is to use the science camera of an ExAO system as a wavefront sensor to perform a fast measurement and correction method to remove these aberrations as soon as they are detected. We have developed the framework for one such method, using the self-coherent camera (SCC), to be applied to ground-based telescopes, called Fast Atmospheric SCC Technique (FAST; Gerard et al., submitted). Our FAST solution requires an optimally designed coronagraph (the SCC FPM) and post-processing algorithm and is in principle able to reach a``raw" contrast of a few times the photon noise limit, continually improving with integration time. In this paper, we present new ongoing work in exploring the manufacturing limitations of the SCC FPM as well as a general framework to implement and optimize a FAST deformable mirror control loop.
\end{abstract}


\section{INTRODUCTION AND SUMMARY OF THE FAST SOLUTION}
\label{sec:intro}  

Direct detection and characterization of exoplanets is a major goal of all future observatories. However, all extreme adaptive optics (ExAO) instruments are still limited in sensitivity by quasi-static wavefront errors, and sometimes additional atmospheric errors. We proposed a new method using the self-coherent camera (SCC\cite{scc_orig,scc_lyot}), to address both of these problems, called the fast atmospheric SCC technique (FAST\cite{gerard18}). The initial FAST paper will hereafter be referred to as G18. The classical SCC technique is reviewed extensively in other papers\cite{scc_orig,scc_lyot,baudoz_psfsubt,mazoyer,mrscc,gerard18}. In this section we review the results from the FAST solution in G18, illustrated in Figure \ref{fig: fast} and outlined below:
\begin{figure}[!h]
\begin{center}
\includegraphics[width=1.0\textwidth]{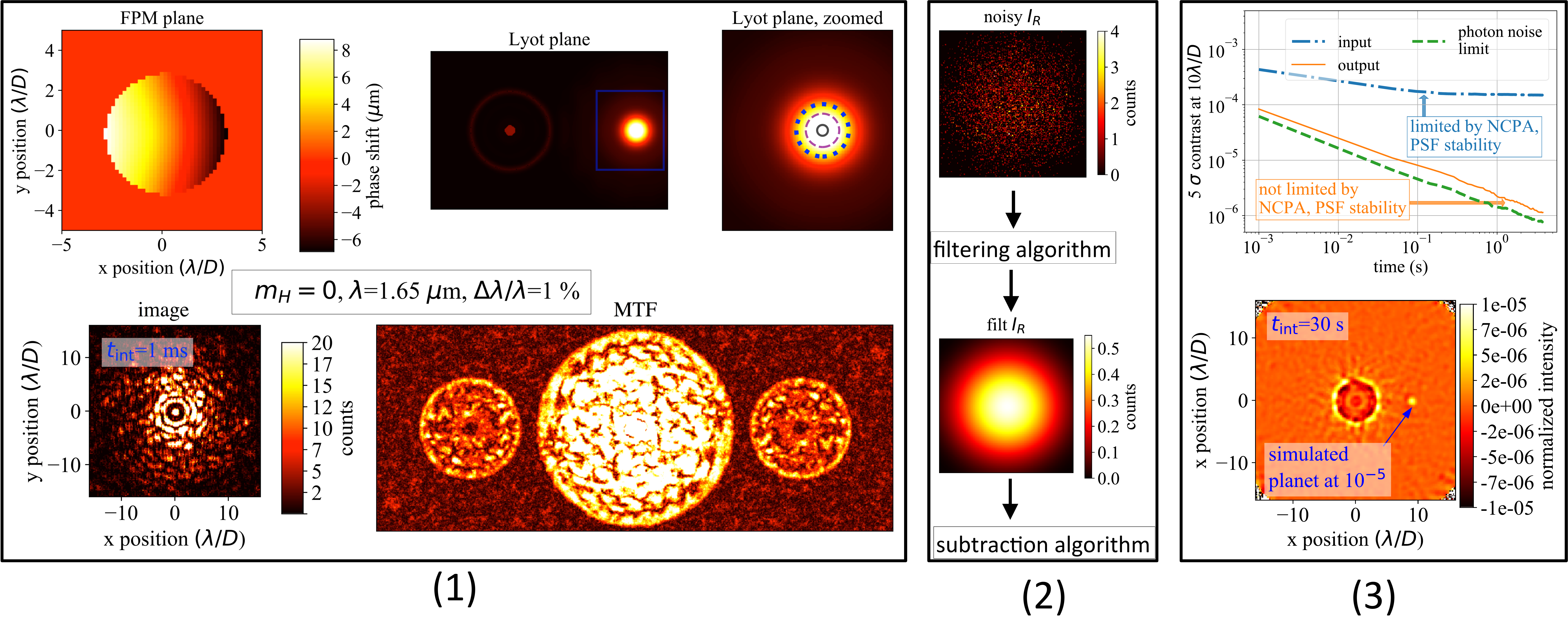}
\caption{An outline of the FAST solution from G18\cite{gerard18}. (1) The SCC FPM (upper left) optimizes the amount of light transmitted through the reference pinhole in the downstream Lyot stop (upper middle and right) to allow detection of fringes for a 1 ms exposure in the further downstream image plane (lower left and right). (2) By simultaneously recording the ``live'' reference pinhole PSF, a reconstruction of the noiseless pinhole PSF allows a subtraction of stellar speckles (but not an exoplanet) close to the photon noise limit. (3) Stacking subtracted images with time continuously follows the photon noise limit, improving contrast simply by integrating longer.}
\end{center}
\label{fig: fast}
\end{figure}
\begin{enumerate}
\item\label{step: fast1} The SCC focal plane mask (FPM; upper left) concentrates light from the point spread function (PSF) core into the Lyot stop pinhole (upper middle and upper right) in order to achieve a sufficient signal-to-noise ratio (SNR) detection of fringes above the photon noise in a 1 ms exposure (lower left and lower right, were MTF is the modulation transfer function). In the upper right, the dashed purple circle shows the theoretical minimum size of this off-axis pupil copy generated from the 6 $\lambda/D$ diameter SCC FPM compared to our measured full width at half maximum (FWHM; dotted blue circle) and SCC pinhole size (solid black circle).
\item A post processing strategy uses the fringe detection to measure and subtract all stellar speckles down to the photon noise limit in exposures on the order of a few milliseconds. $I_R$ is the PSF recorded from the off-axis pinhole in the Lyot stop. Even with the SCC FPM, such a pinhole PSF measurement is dominated by photon noise. A filtering algorithm to remove most this noise allows for an algorithmic subtraction of stellar (but not exoplanet) speckles.
\item Stacking subtracted images over time yields a contrast (see G18 for a formal definition) noise floor that improves proportional to $t^{-0.5}$ ($t$ is time), only slightly above the photon noise limit.
\end{enumerate}
In principle, this approach is no longer limited by either quasi-static aberration or residual atmospheric speckles. More generally, a FAST solution is able to subtract any speckles above the photon noise limit with lifetimes that are longer than the exposure time used in step \ref{step: fast1} above. 

After laying the groundwork for a FAST solution in G18, in this paper we present ongoing work in preparation to manufacture and test the SCC FPM in our new lab, called ``New Earth,'' at the National Research Council of Canada, Herzberg (\S\ref{sec: scc_fpm}) as well as the first simulations using a FAST deformable mirror (DM) control loop (\S\ref{sec: dm_control}). Optimized chromaticity solutions and instrument-specific upgrades will not be addressed in this paper. In this paper we use the same simulation parameters as in G18 unless otherwise noted. 
\section{SCC FPM MANUFACTURING SIMULATIONS}
\label{sec: scc_fpm}
\subsection{Numerical Simulation Parameters}
\label{sec: sim}
For our simulations in this section, we will use $\lambda_0=1.3 \;\mu$m (where $\lambda_0$ is the central wavelength optimized for the SCC FPM), and an accordingly optimized SCC FPM Gaussian FWHM of $4.7 \lambda_0/D$ and amplitude of $2.22\; \mu$m at $\lambda_0$. We sample and integrate over 5 different wavelengths across each bandpass to simulate broadband performance. The simulations in this section do not include photon noise.

Chromatic magnification effects are simulated in the FPM plane by spatially rescaling the phase shift applied by the SCC FPM (i.e., the inner working angle, or IWA, and Gaussian FWHM are scaled by $\lambda/\lambda_0$, where $\lambda$ is a simulated wavelength within the desired bandpass). This effect causes a relatively less-optimal throughput of intensity through the Lyot stop pinhole at wavelengths other than $\lambda_0$. In sections \ref{sec: fringe_visibility} and \ref{sec: rms} we will want to understand how chromatic differences of differential piston between the Lyot stop pupil and pinhole affect fringe visibility. Because this degradation is separate from how magnification with wavelength affects fringe visibility, we do not simulate magnification with wavelength in the detector plane. Therefore, the fringe visibility values will in reality be worse than the results presented in this section at wavelengths further away from $\lambda_0$.
\subsection{Fringe Visibility}
\label{sec: fringe_visibility}
Although our initial goal to do FAST atmospheric wavefront sensing optimized the integrated intensity going through the reference pinhole using the SCC FPM\cite{gerard18}, we instead consider here the metric of fringe visibility. We define fringe visibility as the cumulative flux in the MTF side lobe of an SCC image (see Figure \ref{fig: fast}) divided by the cumulative flux in the central beam of the same MTF, which is denoted as $\Sigma\left(\text{MTF}\times m_1\right)/\Sigma\left(\text{MTF}\times m_2\right)$, where $m_1$ and $m_2$ are binary masks which filter the MTF side lobe and central beam, respectively\cite{gerard18}. This value is similar to our previous metric of power through the pinhole relative to the central pupil; in both cases higher values correspond to higher fringe amplitudes at a single wavelength, where in an optimal regime these numbers are close to unity so a speckle can be measured as soon as it is detected. However, once we start to consider fringe detection over a larger bandpass, these two metrics will differ. Because the phase shift applied on the complex wavefront by the SCC FPM is chromatic, the downstream complex wavefront transmitted through the pinhole will vary as a function of wavelength. Accordingly, the fringe visibility will be lower over a broadband. The intensity pattern in the Lyot plane will only be optimized (i.e., the most concentrated around the pinhole) for a single wavelength, and differential phase offsets between the pupil and the pinhole could cause nulling effects that vary as a function of wavelength. Rather than measuring flux in the Lyot plane, cumulative flux in the MTF is therefore a better metric to measure this type of fringe visibility degredation.

\begin{figure}[!h]
\centering
	\begin{minipage}[b]{0.32\textwidth}
		\begin{center}
		\includegraphics[width=1.0\textwidth]{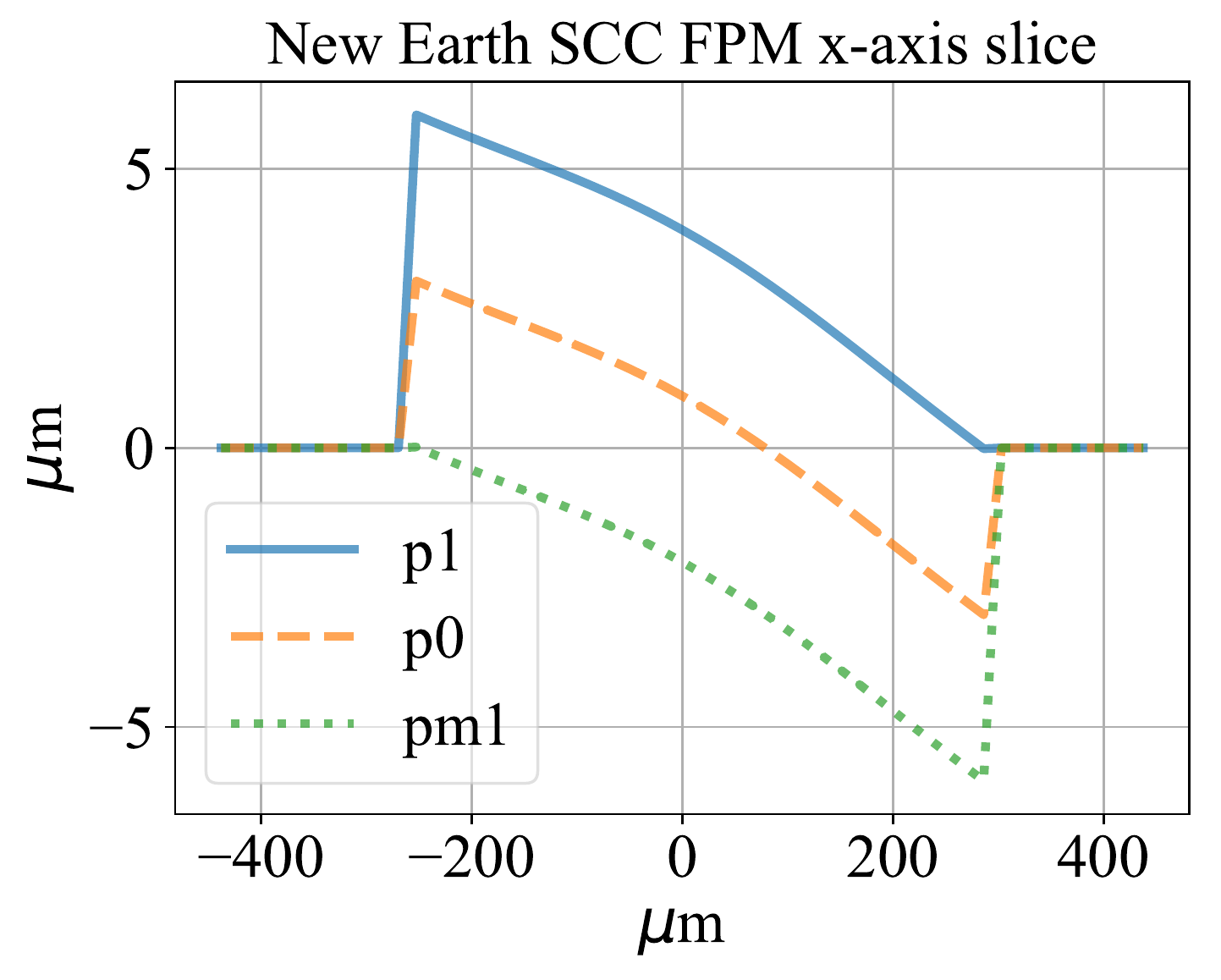}
		(a)
		\end{center}
	\end{minipage}
	\begin{minipage}[b]{0.32\textwidth}
		\begin{center}
		\includegraphics[width=1.0\textwidth]{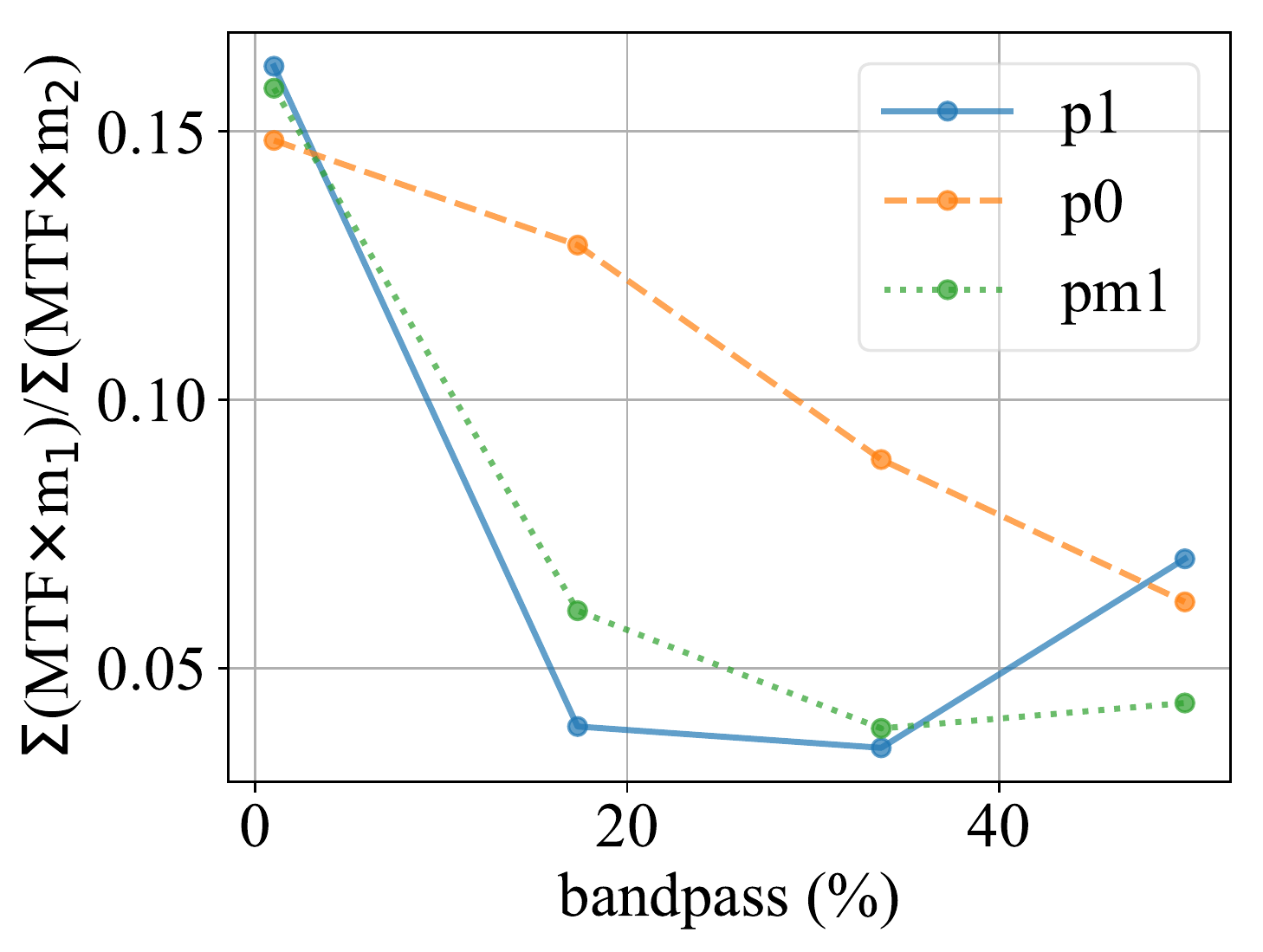}
		(b)
		\end{center}
	\end{minipage}	
	\begin{minipage}[b]{0.32\textwidth}
		\begin{center}
		\includegraphics[width=1.0\textwidth]{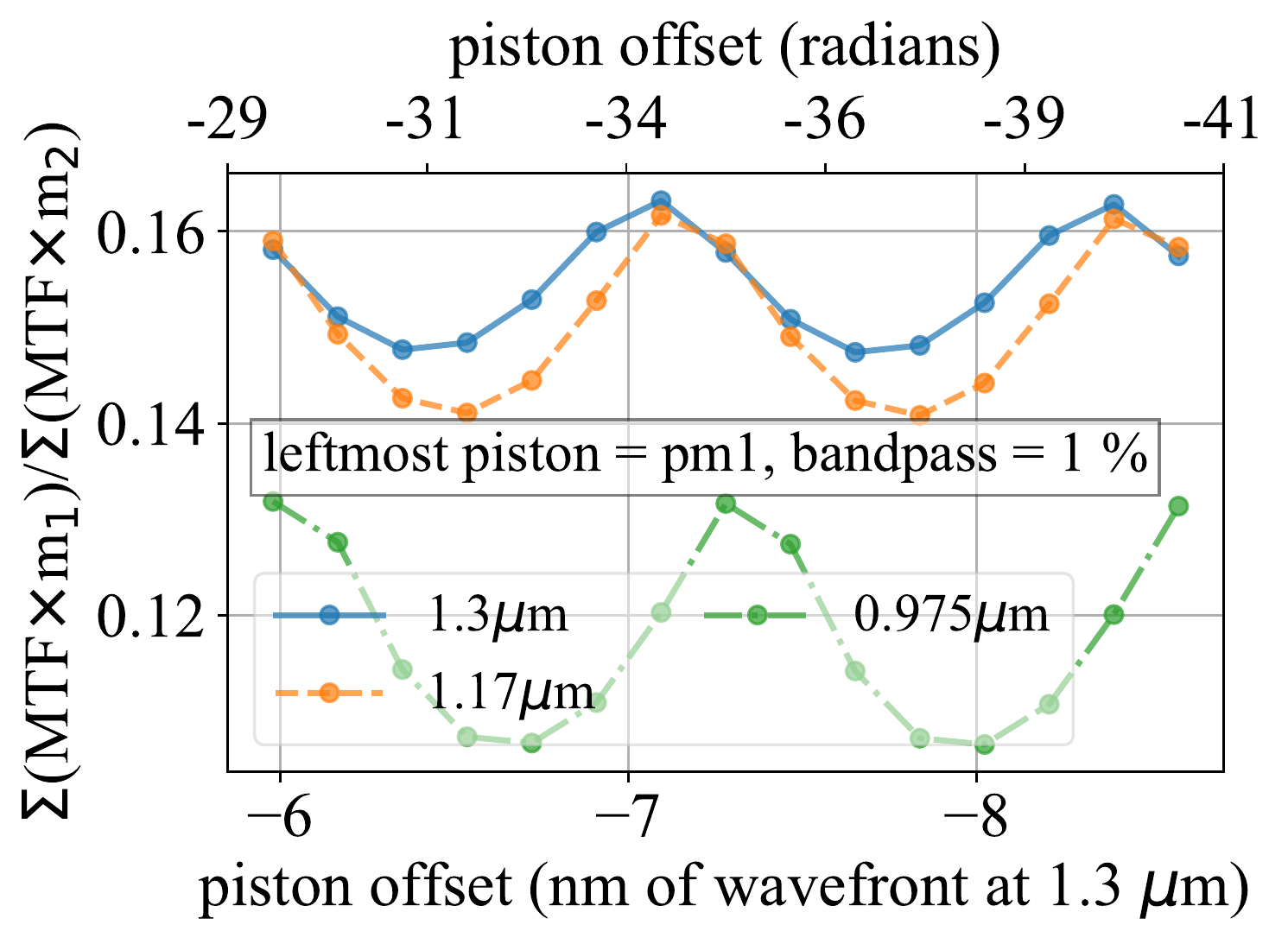}
		(c)
		\end{center}
	\end{minipage}	
\caption{(a) Three different possible reflective SCC FPM designs for our New Earth testbed FPM plane at f/67 and $\lambda_0$=1.3 $\mu$m, showing a slice along the x-axis zoomed in around the central ``bump.'' (b) Fringe visibility vs. bandpass ($\Delta\lambda/\lambda_0\times100$) for the three different designs from (a). (c) Fringe visibility vs. additional negative piston offset for the pm1 design from (a), cycling through 4$\pi$ radians below the pm1 piston value.}
\label{fig: fringe_visibility}
\end{figure}

With these fringe visibility effects and metric in mind, we considered three different possible SCC FPM designs for our New Earth lab, illustrated in Figure \ref{fig: fringe_visibility}a. 
We have actively been working with collaborators and vendors to determine options for manufacturing the SCC FPM from G18. A four step grey scale lithography process---involving (1) fabrication of a greyscale mask, (2) an ultraviolet lithographic process to transfer the grey scale pattern into photoresist , and (3) etching to transfer the photoresist pattern into fused silica, and (4) coating---would require using the pm1 design (S. Thibault, private communication). However, another liquid crystal-based fabrication process could allow using the p0 design (F. Snik, private communication). The p1 and pm1 designs create a chromatic piston phase discrepancy between the pinhole and central pupil in the Lyot plane. Figure \ref{fig: fringe_visibility}b illustrates that this effect significantly decreases the fringe visibility over a large bandpass. Thus the pm1 design will not be a viable option for fabrication of a broadband SCC FPM. However, for a narrow bandpass, which we will consider in this paper, all three designs provide similar fringe visibility. 

Figure \ref{fig: fringe_visibility}c shows the simulation results of how additional piston for a 1\% bandpass affects fringe visibility. Varying the pm1 design by an additional 4$\pi$ radians, constructive and destructive interference cycle the fringe visibility through 2 full periods of oscillation. This curve is shown for three different wavelengths, but for all cases we use the same SCC FPM, optimized for $\lambda_0=$1.3 $\mu$m. Thus, fringe visibility is highest across all piston phase offsets for $\lambda=\lambda_0$, lower for $\lambda=1.17\;\mu$m, and lowest for $\lambda=0.975\;\mu$m (i.e., more light goes through the pinhole closer at wavelengths closer to $\lambda_0$). Additionally, the amplitude of fringe visibility oscillation as a function of continuous piston offset is higher for wavelengths further away from $\lambda_0$, again illustrating that the SCC FPM is an inherently chromatic design.
\subsection{Surface Figure Error}
\label{sec: rms}	
\begin{figure}[!h]
\begin{center}
\includegraphics[width=0.5\textwidth]{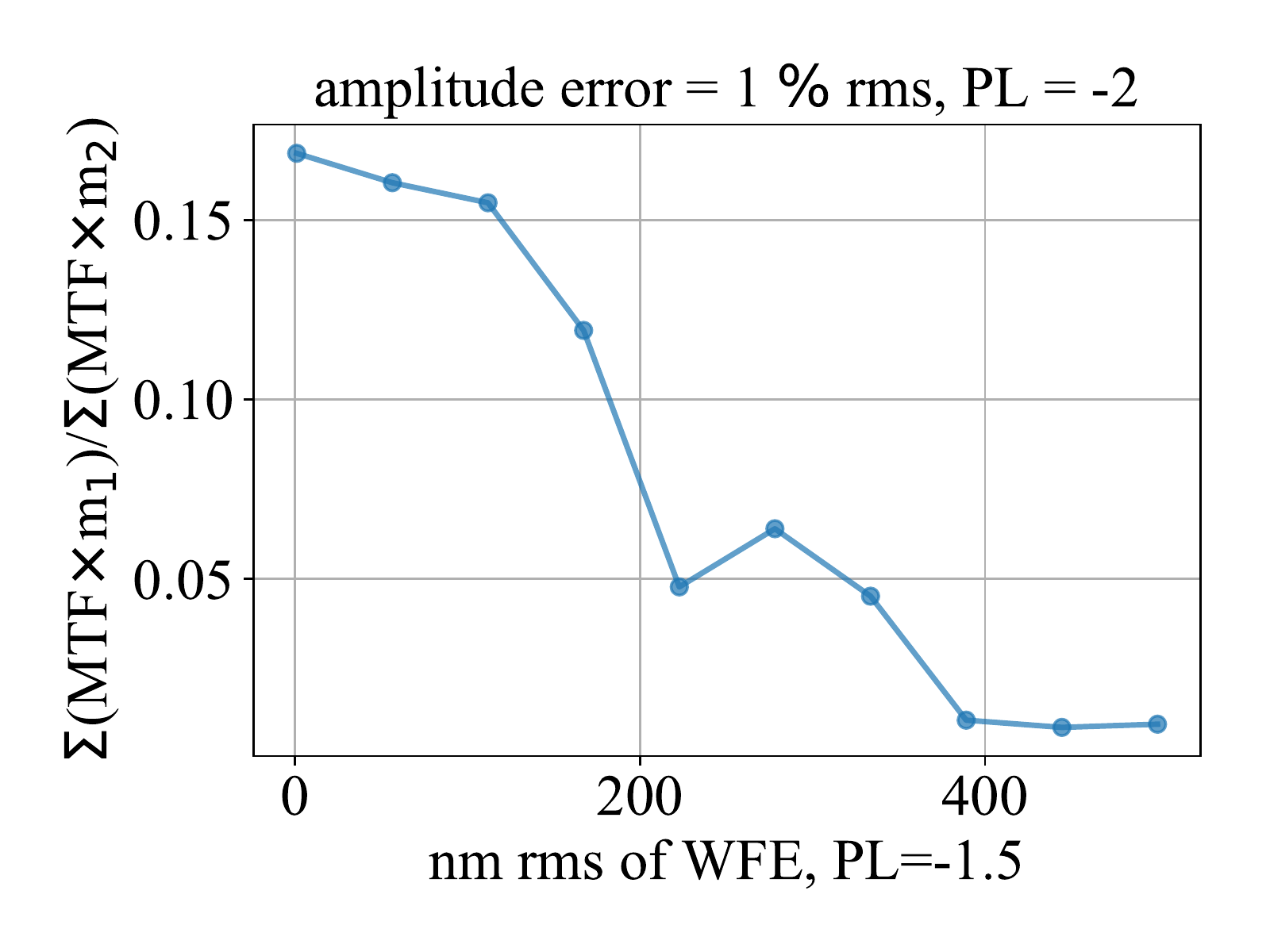}
\caption{Fringe visibility vs. phase wavefront error (WFE) on the surface of the pm1 SCC FPM design, assuming an additional static level of amplitude aberration. Assuming a fringe visibility tolerance requirement of less than about a factor of two from the optimal value, a reflective pm1 SCC FPM design would require less than about 140 nm rms of surface figure error.}
\label{fig: polishing_error}
\end{center}
\end{figure}

Now informed by the results of \S\ref{sec: fringe_visibility}, Figure \ref{fig: polishing_error} shows fringe visibility vs. phase wavefront error (WFE) on the surface of the pm1 SCC FPM design using a 1 \% bandpass centred around $\lambda_0$. We assume a 1\% rms -2 power law (PL) amplitude error.
Imposing a fringe visibility tolerance requirement of better than 0.075 (i.e., better than a factor of two degradation), Figure \ref{fig: polishing_error} illustrates that this would require $\lesssim$200 nm rms of WFE, or $\lesssim$100 nm rms of surface figure error for a reflective SCC FPM design.
\section{FAST DM CONTROL}
\label{sec: dm_control}
\subsection{Numerical Simulation Parameters/Setup}
\label{sec: setup_dm}
Unless specifically stated, all numerical simulation parameters are the same as in G18. As in G18, for static aberration we use $\lambda_0$ = 1.65 $\mu$m, a 1\% bandpass filter (for flux calculation purposes only), a 25 nm rms, -1.5 PL phase screen and 1 \% rms, -2 PL amplitude error, and for residual atmospheric aberration we use a 100 nm rms, -2 PL phase screen. We use a 32 $\times$ 32 actuator DM. Contrast  curves are computed out to 22 $\lambda/D$ by the same definition as in G18 (defined by standard deviation);  only the corners of the DM control region are used to compute contrast beyond 16 $\lambda/D$. Unlike G18, here we use an apodized lyot coronagraph (APLC) design\cite{remi} but still with an optimized SCC FPM; using an apodizer and a 2.5 $\lambda/D$ IWA SCC FPM, we found maximal fringe visibility for a Gaussian aberration with a 2.4 $\lambda/D$ FWHM and 2.8 $\mu$m amplitude. With these parameters the integrated intensity over the pinhole is now about nine times greater than the integrated intensity over the central pupil of the Lyot stop. 

The standard SCC DM control algorithm\cite{baudoz_psfsubt, mazoyer} involves placing a series of sines and cosines on the DM, corresponding to spots in the detector plane at $\lambda/D$ separations from one another throughout the half dark hole (DH). For each recorded sine/cosine image, the standard SCC wavefront sensing algorithm is used to isolate the complex modulation amplitude\cite{scc_orig,scc_lyot,baudoz_psfsubt,mazoyer,mrscc,gerard18}, called $I_-$, generating a ``vectorized'' image of the real and imaginary components of $I_-$ inside the DH. Instead of using the recorded sine/cosine image to calculate $I_-$, we use the difference between a sine/cosine image and an image with a flat DM\cite{baudoz_psfsubt}. This differential approach allows significantly better linearity of the least-squares subtraction; without this approach we require using 30 nm amplitude sines/cosines, whereas after we able to use 5 nm amplitudes for the same spot locations. Although quasi-static speckles can be corrected only for a half DH with a single DM because of the presence of both phase and amplitude aberrations\cite{speckle_nulling}, the same correction will subtract residual atmospheric speckles over a full dark hole because there is $\sim$no atmospheric amplitude aberration (i.e., ignoring effects from scintillation). Then, for an $N\times N$ actuator DM, each vector is multiplied and summed by every other vector to generate a $N^2\times N^2$ covariance matrix. Using a \href{https://docs.scipy.org/doc/numpy-1.14.0/reference/generated/numpy.linalg.pinv.html}{pseudo inverse} to invert the covariance matrix, we tuned the singular value decomposition (SVD) cutoff to optimize contrast after one iteration, finding an optimal value of 0.15 (where a value of 1 would set the inverse covariance matrix to zero). After the pseudo inverse covariance matrix is calculated during a daytime calibration procedure, during on-sky operation it is multiplied by the target image correlation vector to generate least-squares coefficients\cite{loci} for every sine/cosine reference position. Each least squares coefficient is then multiplied by the corresponding entrance pupil sine/cosine phase shift and summed in a linear combination to produce the DM shape, which is then multiplied by negative one (i.e., an optical subtraction), applied to the DM, and propagated through the coronagraph to the detector plane. 

\begin{figure}[!h]
	\begin{minipage}[b]{0.53\textwidth}
		\begin{center}
		\includegraphics[width=1.0\textwidth]{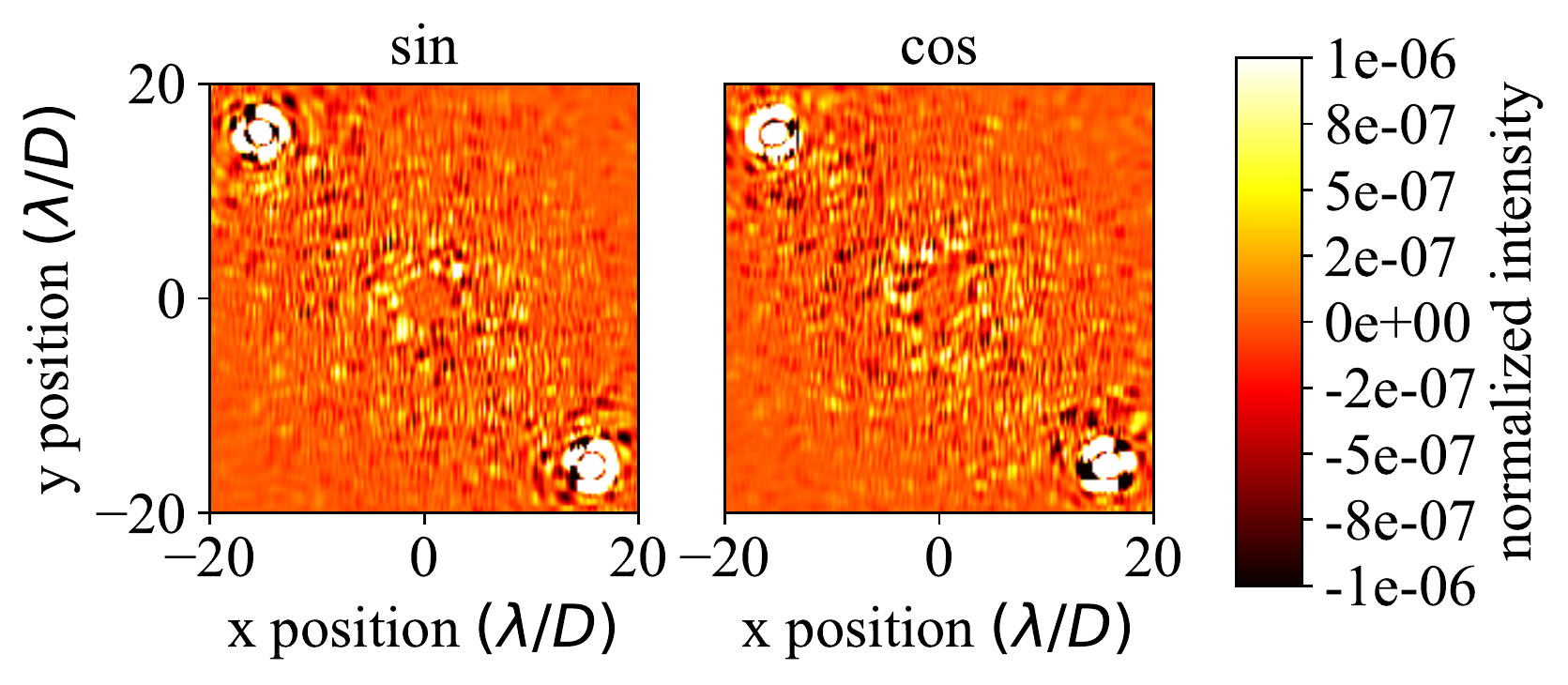}
		(a)
		\end{center}
	\end{minipage}
	\begin{minipage}[b]{0.26\textwidth}
		\begin{center}
		\includegraphics[width=1.0\textwidth]{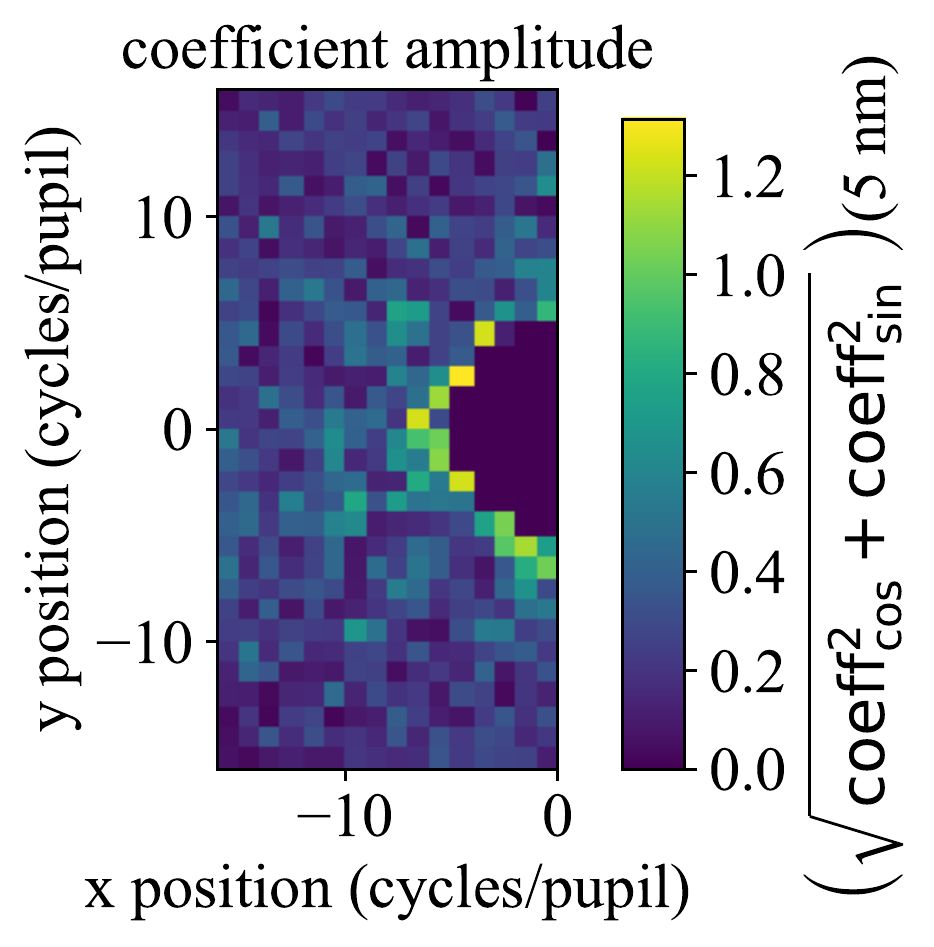}
		(b)
		\label{fig: b}
		\end{center}
	\end{minipage}	
	\begin{minipage}[b]{0.2\textwidth}
		\begin{center}
		\includegraphics[width=1.0\textwidth]{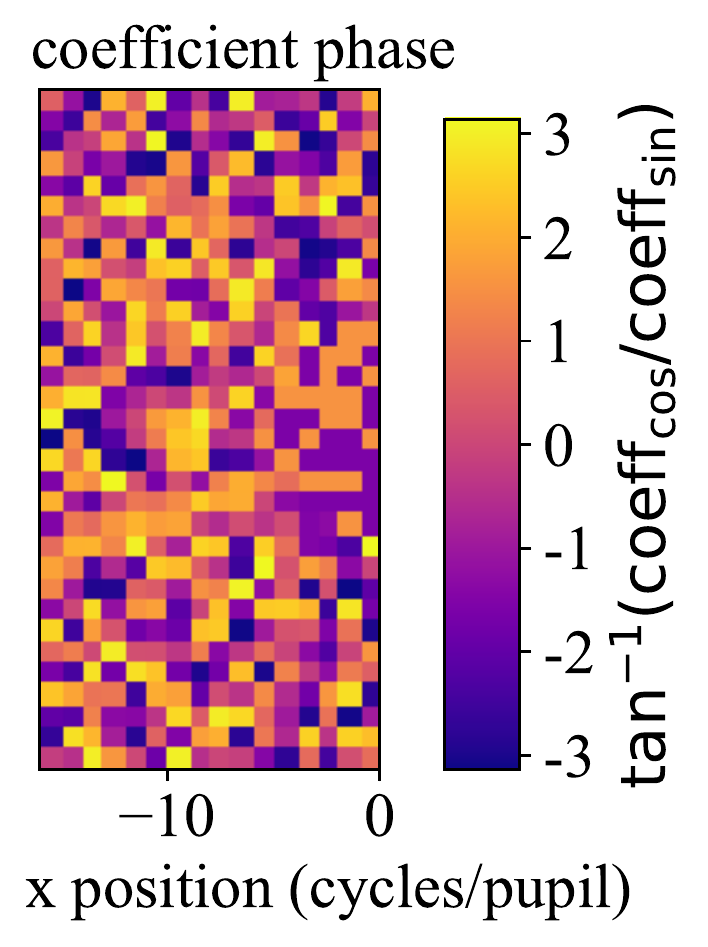}
		(c)
		\label{fig: c}
		\end{center}
	\end{minipage}	
	\begin{minipage}[b]{0.52\textwidth}
		\begin{center}
		\includegraphics[width=1.0\textwidth]{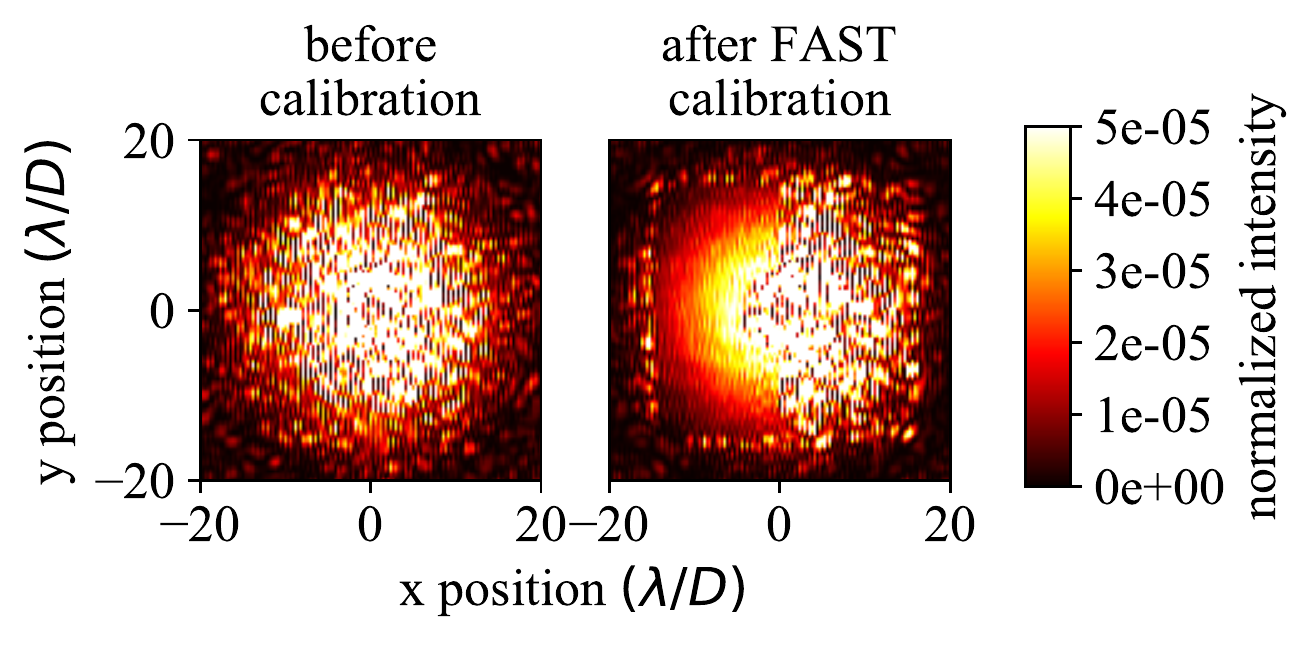}
		(d)
		\label{fig: d}
		\end{center}
	\end{minipage}	
	\begin{minipage}[b]{0.47\textwidth}
		\begin{center}
		\includegraphics[width=1.0\textwidth]{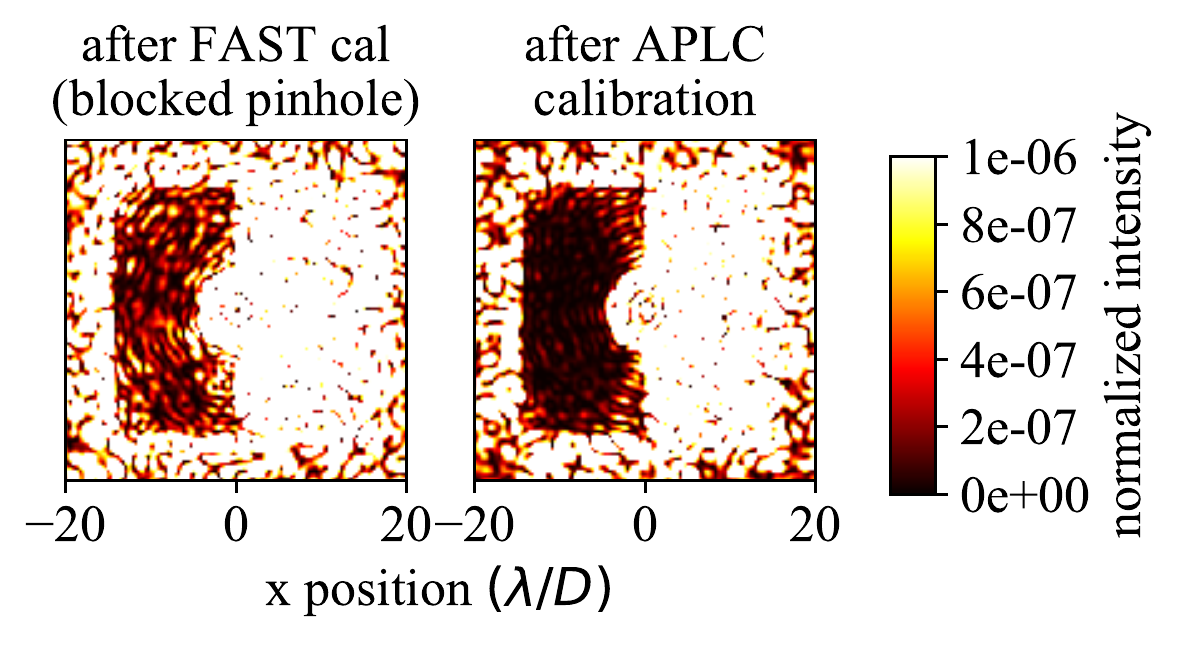}
		(e)
		\label{fig: e}
		\end{center}
	\end{minipage}	
	\begin{center}
	\begin{minipage}[b]{0.8\textwidth}
		\begin{center}
		\includegraphics[width=1.0\textwidth]{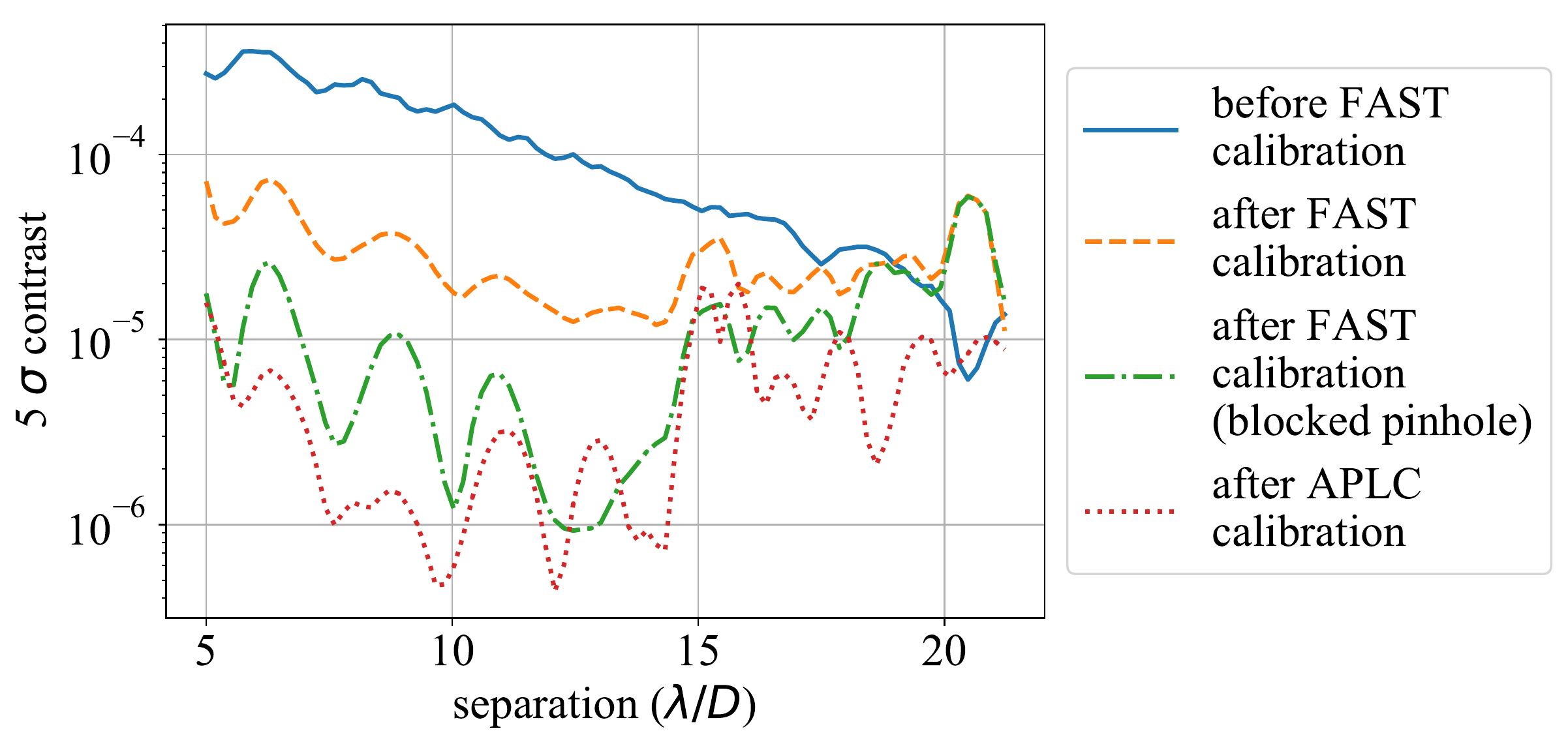}
		(f)
		\label{fig: f}
		\end{center}
	\end{minipage}	
	\end{center}
\caption{(a) The difference of a recorded SCC image with a DM sine and cosine spot (each at the same spatial frequency) and a recorded SCC image with no sine spot, where otherwise both images have the same static wavefront (25 nm rms, 1 \% amplitude rms). (b) and (c) The least-squares coefficients in phase and amplitude, respectively, generated from an uncalibrated target image to be applied to the 5 nm reference sine/cosine DM spots, shown as an amplitude and phase, respectively. (d) An uncorrected image with the same static aberration use to generate (a), (b), and (c), labeled ``before calibration,'' and a calibrated image after applying a FAST correction with one DM, labeled ``after FAST calibration.'' (e) left: the same DM commands used to generate the image in the right pannel of (d) but with the reference pinhole blocked. right: a calibrated correction with one DM using the APLC\cite{remi} (an amplitude FPM instead of the SCC FPM, but still with the same 2.5 $\lambda/D$ IWA); the same calibration procedure used to generate the right panel of (d) is followed. (f) contrast curves (as defined in G18) for images in (d) and (e). }
\label{fig: dm_setup}
\end{figure}

We also modified some steps of the standard calibration procedure described above that required additional tuning to optimize the final contrast in a calibrated image:
\begin{enumerate}
\item Sine/cosine spots are placed at $\lambda/D$ intervals but at 0.5 $\lambda/D$ offset from the center of the star. Thus, in x and y offsets from the optical axis in the detector plane, the spot positions are between 0.5 and 15.5 $\lambda/D$ at 1 $\lambda/D$ increments, still yielding $N^2$ reference spots and a $N^2\times N^2$ covariance matrix.
\item We found better linearity and contrast of the DM correction by setting the reference sines/cosines with radial separation $\leq 5 \lambda/D$ equal to zero. This is expected for separations $\leq 2.5 \lambda/D$ (the IWA), for which any exoplanet would be blocked by the FPM; however, we also found that not subtracting speckles pinned to the bright diffraction rings between 2.5 and 5 $\lambda/D$ improved contrast and linearity of the DM correction. Thus, the effective IWA in our simulations is 5 $\lambda/D$.
\end{enumerate}
Figure \ref{fig: dm_setup} illustrates the process and results of our calibration procedure after including the additional modifications described above. 

Figure \ref{fig: dm_setup} clearly shows that the pinhole PSF of the SCC FPM limits achievable contrast. Our current design concentrates too much light through the reference pinhole; because there is nine times more flux going through the pinhole than the central pupil, the pinhole PSF is now above the speckle noise floor and limiting the achievable contrast. The ``raw'' 5$\sigma$ contrast in Fig. \ref{fig: dm_setup} d converges to $7\times10^{-5}$ over the DH, a contrast improvement by a factor of 3. When the same DM correction from Fig. \ref{fig: dm_setup} is applied with the pinhole blocked in \ref{fig: dm_setup} e, the DH contrast instead improves by a factor of 16. Even if the pinhole PSF was optimally matched to the speckle amplitudes, the same limiting physical principle would ultimately apply: assuming the zeroth order diffraction noise floor is sufficiently below the second order speckle halo term\cite{perrin_psf} (i.e., using a coronagraph optimized to suppress diffraction), a calibrated image with the SCC FPM will ultimately be limited by the pinhole PSF, not the speckle halo. At this time we do not address this problem; in this paper will only consider 2 second exposures to demonstrate the basic framework for FAST DM control, and thus we do not expect to reach better than a factor of 3 contrast improvement. 

Fig. \ref{fig: dm_setup} e illustrates a second problem: compared to a 2.5 $\lambda/D$ IWA amplitude FPM, the 2.5 $\lambda/D$ IWA SCC FPM cannot reach the same contrast levels. Calibrated speckles in the DH of the right panel of Fig. \ref{fig: dm_setup} e improve by a factor of 1.8 compared to left panel. Because we do not simulate photon or detector noise here (i.e., assuming a daytime calibration using a bright internal source), the physical origin of this discrepant noise floor must arise from the difference between the wavefronts transmitted through the pinhole when using an amplitude FPM vs. SCC FPM. At this time we do not address this discrepancy and simply illustrate the potential limitation, although the contrast curve in Fig. \ref{fig: dm_setup} f illustrates that this problem is a second order effect compared to the limiting pinhole PSF amplitude effect discussed above.
\subsection{Numerical Simulations}
\label{sec: sims_dm}
\subsubsection{Framework}
\label{sec: framework}
We ran simulations of a fast DM control loop using the optimized calibration procedure described in \S\ref{sec: setup_dm}. Temporal power spectral density (PSD) plots are estimated by computing a periodogram of the time-series data\cite{numerical_methods}. To remove low frequency under-sampled data we first apply a linear detrending procedure to the time series and then implement a \href{https://www.docs.scipy.org/doc/scipy-0.14.0/reference/generated/scipy.signal.welch.html}{welch} routine\cite{welch}, using 256 segments and a Hanning window to suppress noise in the PSD. On-sky images use a $m_H=0$ star, 250 Hz frame rate ($T_s$ = 4 ms), and a 1 millisecond delay ($\tau$ = 1 ms). This delay is not unrealistic. For a typical CPU, the speed for a fast Fourier transform (FFT) of a real two dimensional image at single precision is about 15,000 mflops\cite{mflops}. For a 256$\times$256 image (a beam ratio of 8 if sampling only the DM control region), this would take 175 $\mu$s per FFT\cite{fftspeed}. Two FFTs to compute $I_-$ would take 350 $\mu$s. Combining the 1024$\times$1024 (rows$\times$columns) inverse covariance matrix, 1024$\times$3610 reference image matrix, and 3610$\times$1 target image vector (the latter two are typically dotted to produce the target image correlation vector) into a single matrix operation produces a $(1024\times3610)\cdot(3610\times1)$ matrix multiplication. One Narrow Field InfraRed Adaptive Optics System (NFIRAOS) server processing one laser guide star wavefront sensor (WFS) multiplies a matrix of 7000$\times$5400 by a 5400$\times$1 vector in 500 $\mu$s\cite{nfiraos_rtc}, and so our FAST approach should be less than this amount. Finally, the FFT cannot start before the last pixel is received, so adding a few hundred microseconds to read out the image yields (350 $\mu$s) + (less than 500 $\mu$s) + (about 200 $\mu$s) = (less than about 1.05 ms).

We tested two different methods of temporal DM control using an integrator controller: a constant gain and a optimized modal gain approach\cite{gendron}. We use a standard model for the open loop, closed loop, and noise transfer functions for an integral controller\cite{ao_book, jp}, hereafter respectively denoted as $H_\text{OL}$, $H_\text{rej}$ and $H_n$, which are a function of the gain ($g$), $T_s$, and $\tau$. For the constant gain controller we will demonstrate results for both $g=1.1$ and $g=0.2$. For $T_s=4$ ms, $\tau=1$ ms, and $g=1.1$, $H_\text{OL}$ has a 45$^\circ$ phase margin, a common tolerance requirement to provide the best balance of temporal rejection and system stability for an unoptimized controller\cite{jp}. However, the optimal gain is ultimately governed by the balance of both high atmospheric rejection and low wavefront sensor (WFS) noise amplification, or
\begin{equation}
\text{min}\left\{\text{WFE}_{i}(g=g_\text{opt})\right\}=\text{min}\left\{\sqrt{\int_0^{f_n}df\; \text{PSD}^{'}_i \; |H_\text{rej}(T_s,\tau,g)|^2+\int_0^{f_n}df\; \text{PSD}_{(n,\; i)} \; |H_n(T_s,\tau,g)|^2}\right\},
\label{eq: hrej_hn}
\end{equation}
where ``min\{\}'' is a minimization operator over the integrator gain, $i$ represents a single Fourier mode of the wavefront, $f_n=1/(2 T_s)$ is the Nyquist frequency of the system frame rate, the open loop PSD of pure atmospheric turbulence is PSD$^{'}_i$, PSD$_{(n, i)}$ is the open loop PSD of pure WFS noise (i.e., flat at all temporal frequencies), and $g_\text{opt}$ is the optimal gain governed by the above equation. However, because we do not have direct access to either PSD$^{'}_{i}$ or PSD$_{(n, i)}$, it can be shown\cite{noisy_psd} that instead equation \ref{eq: hrej_hn} is analogous to 
\begin{equation}
\text{min}\left\{\text{WFE}_{i}(g=g_\text{opt})\right\}=\text{min}\left\{\sqrt{\int_0^{f_n}df\; \text{PSD}_i \; |H_\text{rej}(T_s,\tau,g)|^2}\right\},
\label{eq: hrej}
\end{equation}
where PSD$_i$ is now the PSD from noisy time series measurements of open loop coefficients at a single Fourier mode, measuring the temporal statistics of both atmospheric turbulence and WFS noise. With this framework, our implementation of FAST modal gain optimization is outlined below and in Figure \ref{fig: gopt}:
\begin{figure}[!h]
	\begin{minipage}[b]{0.40\textwidth}
		\begin{center}
		\includegraphics[width=1.0\textwidth]{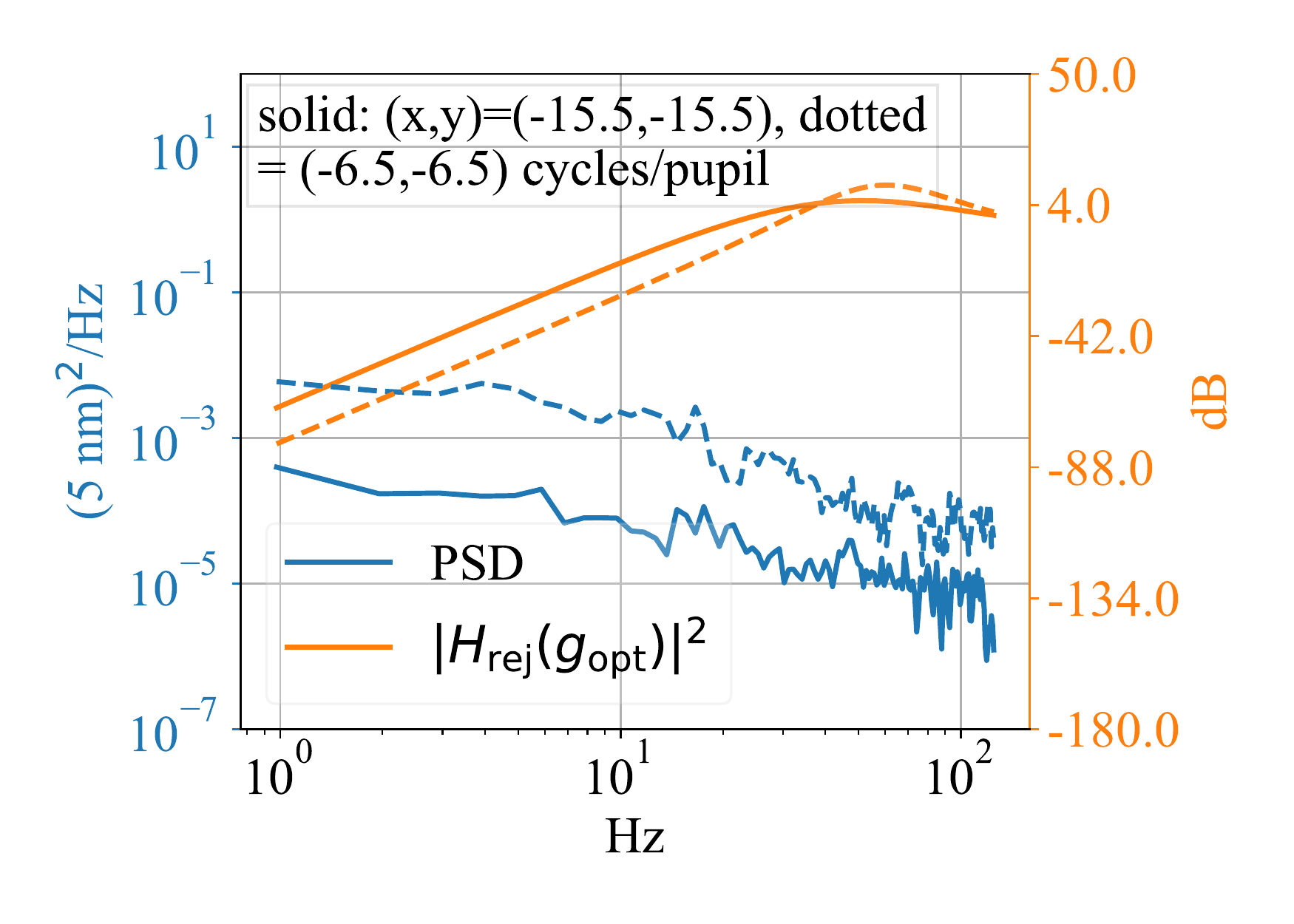}
		(a)
		\end{center}
	\end{minipage}
	\begin{minipage}[b]{0.29\textwidth}
		\begin{center}
		\includegraphics[width=1.0\textwidth]{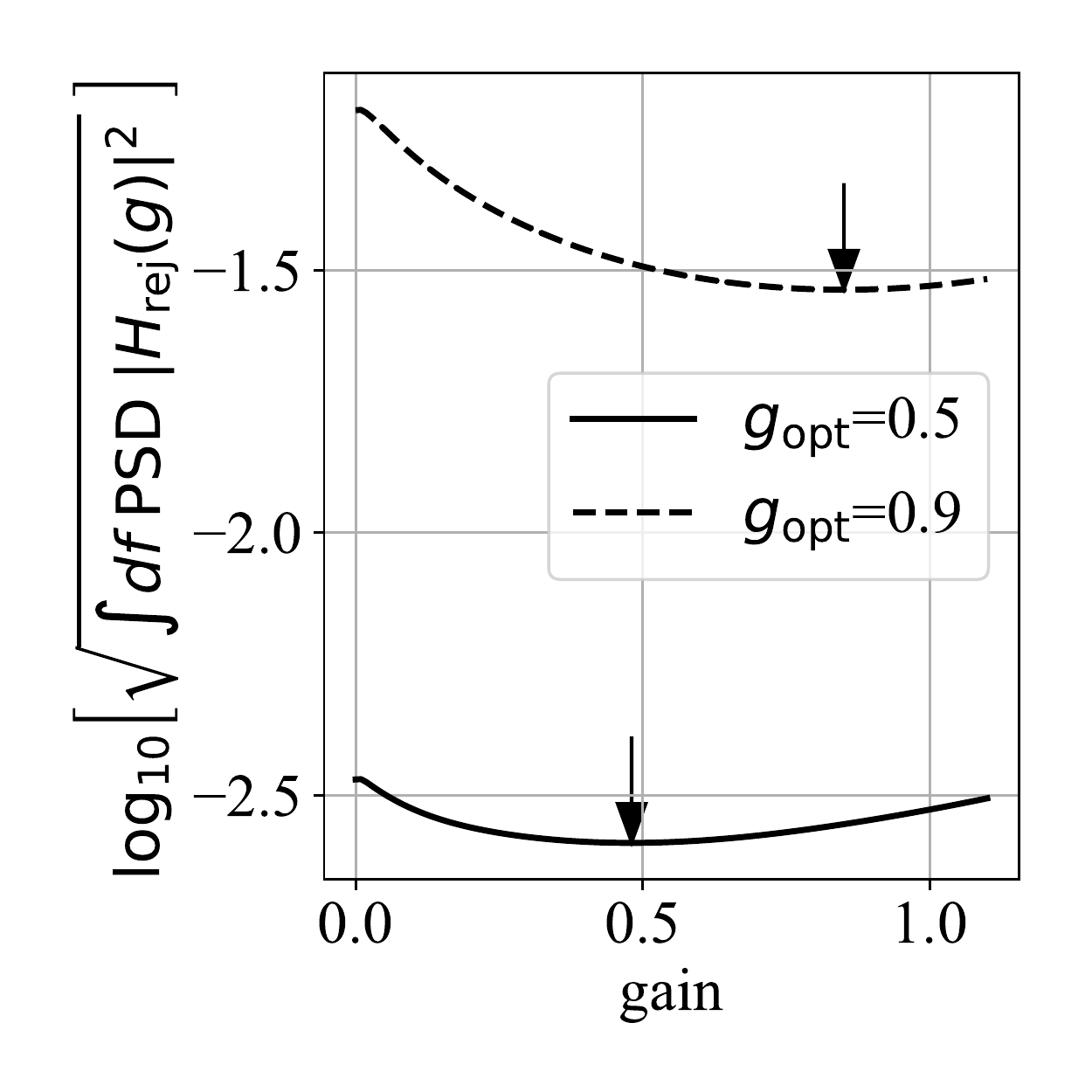}
		(b)
		\label{fig: b}
		\end{center}
	\end{minipage}	
	\begin{minipage}[b]{0.31\textwidth}
		\begin{center}
		\includegraphics[width=1.0\textwidth]{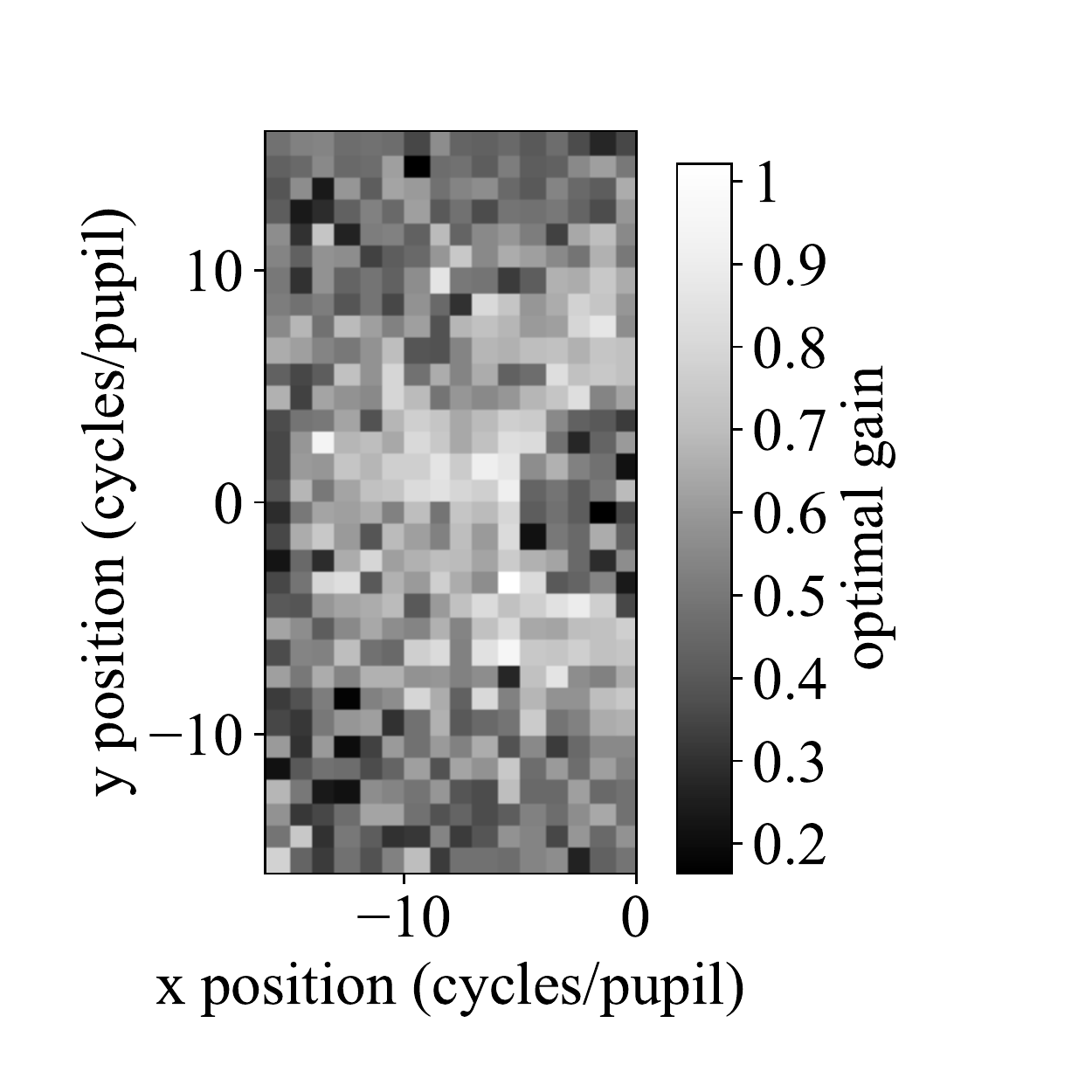}
		(c)
		\label{fig: c}
		\end{center}
	\end{minipage}	
\caption{An outline of the modal gain optimization procedure used for FAST DM control. (a) For two different Fourier modes, plots of both open loop temporal PSDs and the square modulus of the rejection transfer function for the optimal gain found in (b). (b) Gain optimization curves for the same two Fourier modes from (a), governed by equation \ref{eq: hrej}; the optimal gain for each mode is shown in the legend and labeled by an arrow. (c) the optimal gain for every Fourier mode in the DH.}
\label{fig: gopt}
\end{figure}
\begin{enumerate}
\item\label{ref: measure} Measure open loop SCC DM coefficients to sufficiently sample the temporal statistics. We used 2000 frames (8 seconds).
\item\label{step: mode} For the coefficient amplitude of a single Fourier mode (i.e., $\sqrt{\text{coeff}_{\text{cos, }i}^2+\text{coeff}_{\text{sin, }i}^2}$, or a single pixel in Figure \ref{fig: dm_setup} b),
	\begin{enumerate}
	\item construct open loop PSDs (Figure \ref{fig: gopt} a) and then
	\item\label{step: gopt} find the optimal integrator gain using equation \ref{eq: hrej} (Figure \ref{fig: gopt} a and b). 
	\end{enumerate}
\item Repeat step \ref{step: mode} for every Fourier mode to produce an optimal gain map (Figure \ref{fig: gopt} c). 
\item After the open loop measurements from step \ref{ref: measure} are finished recording, apply the optimal gain map from step \ref{step: gopt} in closed loop.
\end{enumerate}
If the rejection transfer function in step \ref{step: gopt} above is not known, it can instead be estimated empirically via\cite{measure_rtf}
\begin{equation}
|\tilde{H}_\text{rej}(g_\text{CL})|^2=\text{PSD}_\text{(i, CL)}\left(g_\text{CL}\right)/\text{PSD}_i,
\label{eq: measure_hrej}
\end{equation}
where $\text{PSD}_\text{(i, CL)}$ is the closed loop PSD for mode $i$, $g_\text{CL}$ is the gain used during those closed loop measurements (not necessarily $g_\text{opt}$), and $\tilde{H}_\text{rej}$ is the estimated rejection transfer function that can replace $H_\text{rej}$ in equation \ref{eq: hrej}. 

Figure \ref{fig: gopt} illustrates the potential advantage of gain optimization. The optimal gain map in Fig. \ref{fig: gopt} c clearly shows a dependence on Fourier mode, illustrating that speckles at the edge of the DH are detected at a relatively lower SNR than speckles closer in, and so the optimal gains are adjusted accordingly. As expected, the PSDs in Fig. \ref{fig: gopt} a show a power law at low temporal frequencies (from atmospheric speckles) and a flat spectrum at high temporal frequencies (from photon noise). The corresponding optimal rejection transfer function balances the maximal rejection of atmospheric speckles at low temporal frequencies and minimal amplification of photon noise at high temporal frequencies. For the Fourier mode closer to the star compared to the mode further away, an atmospheric speckle is brighter (causing a relatively higher PSD amplitude at low temporal frequencies), but as a result the photon noise is also larger (causing a relatively higher PSD amplitude at high temporal frequencies). However, because the SNR of an atmospheric speckle will increase proportional to $t^{0.5}$ (when $t$ is less than the speckle lifetime), as expected the optimal gain is still higher for the Fourier mode closer to the star.

It is also important to note here that equation \ref{eq: hrej} generalizes to any fast wavefront sensing algorithm and any controller. As the temporal statistics change, $H_\text{rej}$ can change (ideally matched to the inverse of the PSD) but still remain optimized by equation \ref{eq: hrej}. This optimization procedure may be over multiple free parameters for more sophisticated controllers (e.g., using a leak controller\cite{dessenne99, measure_rtf}). Although slower ``drifting'' effects from quasi-static aberration and/or diffraction may not be properly sampled to meet the requirement of equation \ref{eq: hrej}, the detrending of open loop temporal PSDs (\S\ref{sec: framework}) should thus not affect the temporal rejection of atmospheric speckles.
\subsubsection{Results}
\label{sec: results}
\begin{figure}[!h]
	\begin{minipage}[b]{0.59\textwidth}
		\begin{center}
		\includegraphics[width=1.0\textwidth]{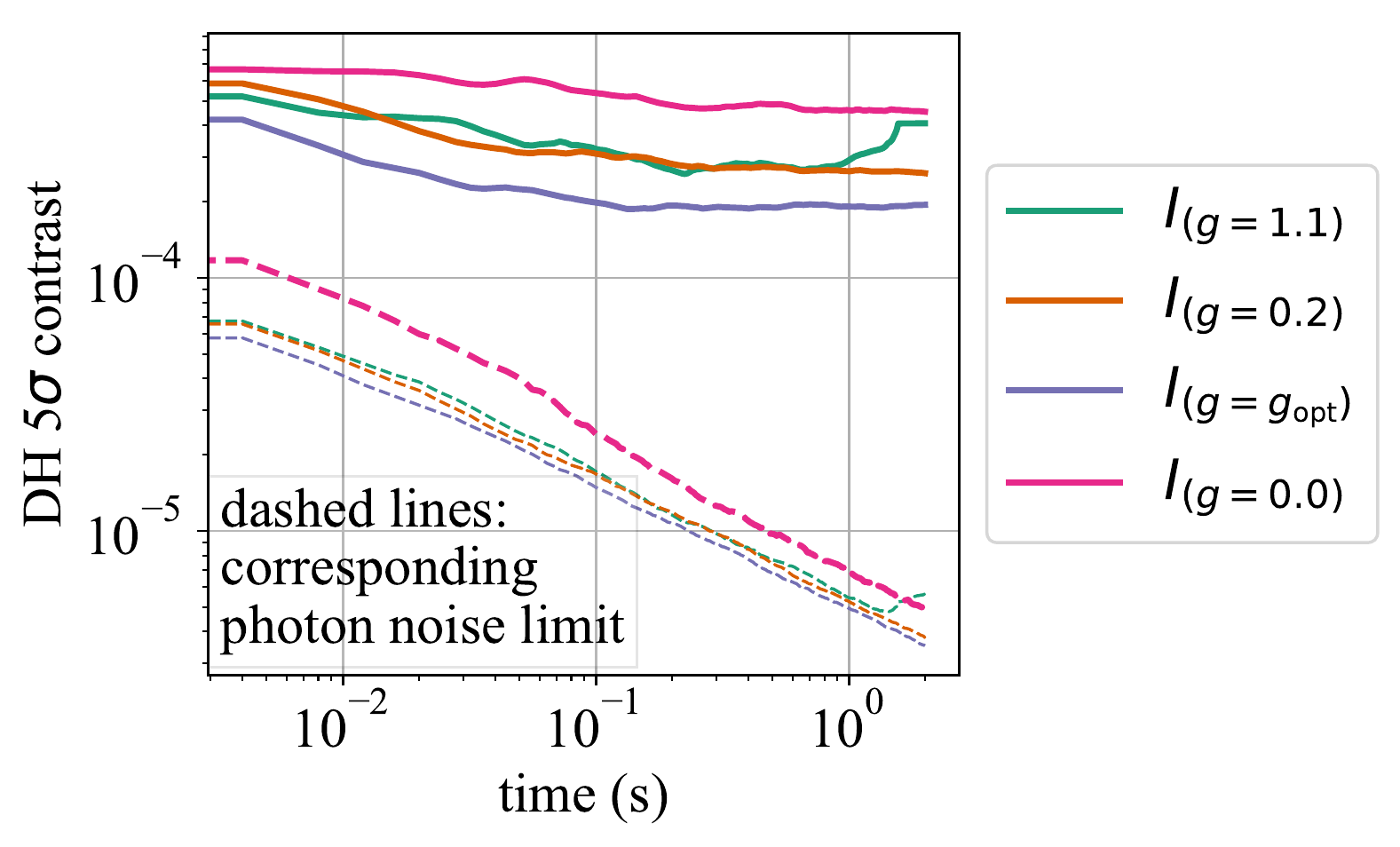}
		\end{center}
	\end{minipage}
	\begin{minipage}[b]{0.4\textwidth}
		\begin{center}
		\includegraphics[width=1.0\textwidth]{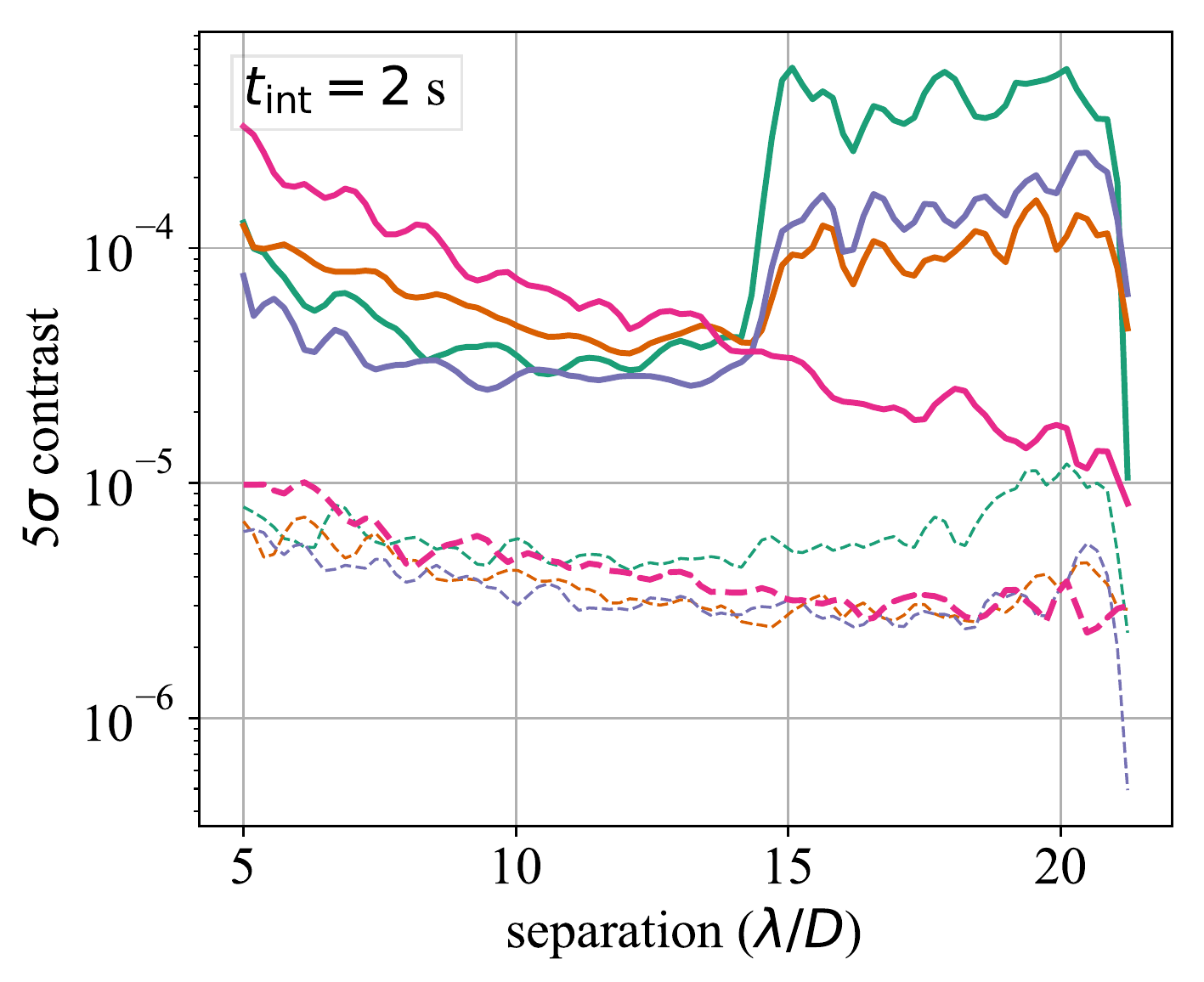}
		\end{center}
	\end{minipage}	
	\begin{center}
	\begin{minipage}[b]{1.0\textwidth}
		\includegraphics[width=1.0\textwidth]{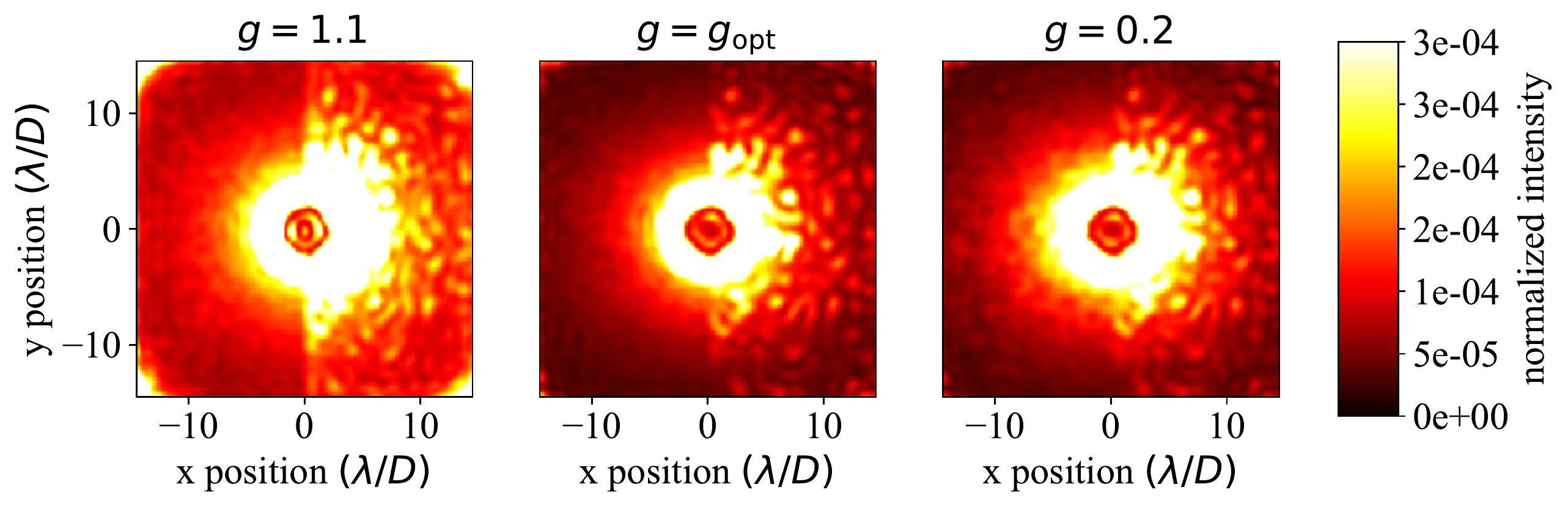}
	\end{minipage}	
	\end{center}
\caption{Results from our two second FAST DM control simulations. upper left: 5$\sigma$ contrast over the full DH vs. time for a low, high, and optimized gain, compared to the respective photon noise limits and the algorithmic photon noise limit from G18. upper right: contrast curves after two seconds of integration time for the images from the upper left panel. lower left, middle and right: images corresponding to the upper right contrast curves for the high, optimized, and low gain simulations, respectively.}
\label{fig: gopt_results}
\end{figure}

Figure \ref{fig: gopt_results} shows the results after two seconds of integration time in closed loop of both a low and high constant gain compared to modal gain optimization. Open loop images (i.e., with no FAST DM control) are denoted as $I_{g=0.0}$. The photon noise limit for each simulation is shown as a dashed line; the open loop photon noise limit (the pink dashed line) is represented by the limit for a perfect algorithmic subtraction from G18. Figure \ref{fig: gopt_results} illustrates both the general potential advantages of optical subtraction over algorithmic subtraction and well as the advantage of gain optimization over a constant gain.

First, as expected the photon noise limits of all optically subtracted images lie below the algorithmic photon noise limit. The clear advantage of optical subtraction is thus when the contrast can reach below the algorithmic-only subtraction limit. If this can be achieved, deeper contrasts will be reached faster. E.g., if optical subtraction is a factor of two below the open loop photon noise limit, reaching a 5$\sigma$contrast of $10^{-8}$ at 10 $\lambda/D$ will take 84 hours of telescope time for the algorithmic-only subtraction (from G18) but only 21 hours for the aforementioned optical subtraction. Although the actual contrast curves for the optically subtracted images are still far from the algorithmic photon noise limit, additional modifications to FAST DM control will bring optical subtraction closer to (and ideally below) this limit, including addressing the static pinhole PSF-related problems discussed in \S\ref{sec: setup_dm} and more sophisticated controllers. For the former problem, we showed in \S\ref{sec: setup_dm} that Additionally, algorithmic subtraction methods from G18 are fully compatible with optically subtracted images, allowing a two-stage (optical$+$algorithmic) subtraction procedure to help close the gap with the algorithmic-only photon noise limit (e.g., a simultaneous pinhole PSF measurement as in G18 \textit{would} allow continuous algorithmic subtraction down to the corresponding optical photon noise limit).

Second, independent of being able to reach below the algorithmic photon noise limit or not, modal gain optimization clearly outperforms both a low and high constant gain. Compared to the results for modal gain optimization, the upper right panel of Fig. \ref{fig: gopt_results} illustrates that a high gain reaches a similar contrast closer to the star but is worse further from the star, while a low gain reaches a similar contrast further from the star but is worse closer to the star, explaining the results over the full DH in the upper left panel. Along with the images in the lower panel, these results clearly illustrate that a higher gain is required closer to the star (where speckles are detected with a higher SNR) for optimal speckle rejection, while a lower gain is required further away from the star (where speckles are detected with a lower SNR) to prevent noise amplification, consistent with the computed optimal gain map Fig. \ref{fig: gopt} c. With that said, although the gain optimizer performs better overall compared to either constant gain, noise is clearly still amplified compared to open loop images beyond about 16$\lambda/D$. Thus, additional approaches and/or modifications to this setup will ultimately be required to minimize noise amplification, such as a smaller pinhole and/or an anti-aliasing spatial filter\cite{sf}.

\section{CONCLUSION}
\label{sec: conclusion}
In this paper we build on the work of G18 to understand the tolerance requirements for manufacturing the SCC FPM and to simulate FAST DM control. We find that sufficient fringe visibility can be achieved for a 1 \% bandpass centered at 1.3 microns if the SCC FPM surface figure is manufactured to better than about 200 nm rms wavefront error. Using a FAST DM control scheme, we also demonstrate a clear advantage of modal gain optimization over a classical constant-gain integral controller.

\acknowledgments 
We gratefully acknowledge research support of the Natural Sciences and Engineering Council of Canada through the Postgraduate Scholarships-Doctoral program, Technologies for Exo-Planetary Science Collaborative Research and Training Experience program, and Discovery Grants Program.

\bibliography{report} 
\bibliographystyle{spiebib} 

\end{document}